\documentstyle[12pt,a4]{article}

\input{epsf.tex}

\begin{document}

\begin{center}
{\bf PROBLEMS OF EXPERIMENTAL DETERMINATION OF PARAMETERS OF NUCLEUS AND
APPLICABILITY OF THE BOHR-MOTTELSON HYPOTHESIS }\\
\end{center}\begin{center}
{ A.M. Sukhovoj, V.A. Khitrov}\\
\end{center}\begin{center}
{\it Frank Laboratory of Neutron Physics, Joint Institute
for Nuclear Research, 141980, Dubna, Russia}\\
\end{center}

Population of a number of excited levels of $^{51}$V and $^{57}$Fe has been determined
from the data of the ENDSF file up to the excitation energy of about 7
and 5.2 MeV, respectively.
It cannot be reproduced in the region of their maximal energies in the
framework of assumption on the independence of partial radiative widths
on the structure of decaying level and excited one, at least,
lower than $0.5B_n$.
Therefore, it is impossible to estimate the actual degree of reliability
of data on the level density and radiative widths of cascade gamma-transitions
in light spherical nuclei obtained from the spectrum analysis of nuclear
reactions, which do not take into account this circumstance.

\section{Introduction}

Large level density at the excitation energy higher than 0.5Bn does not allow
one to obtain experimental data on the level density and partial widths of
gamma transitions by means of the classical nuclear spectroscopy.
These parameters may only be found here from the solution of inverse problem
of the mathematical analysis -- the determination of parameters x, y, z...
of the function S from its values at the relation prescribed a priori
S=f(x,y,z...). In general case the function f(x,y,z...) is determined with
some uncertainty, for example, using an erroneous hypothesis instead of the
lacking experimental information.
In such case the presence and value of the error in the function f(x,y,z...)
may only be determined from experiment, for example, using a hypothesis on
the independence of values of partial radiative widths of $\Gamma$
gamma-transitions of the assigned type and energy on the nucleus excitation
energy.
In the general form this hypothesis has been formulated by Bohr and Mottelson
[1] as an assumption on the independence of the interaction cross section of
the nuclear reaction product on the nucleus excitation energy $U$.
In other words, it does not depend on the structure of nucleus at the energy
$U$, and up to now [2] it is used at the determination of level density from
spectra of evaporative nucleons without any verification.
The same hypothesis in the form of ``the Axel-Brink hypothesis" serves as
a basis for the technique of joint extraction of $\rho$ and $\Gamma$ 
from gamma spectra of the reaction ($^3$He, $\alpha$) in technique [3].

The necessity of introducing any ``zero" hypothesis (of the type [1])
is determined by the lack of necessary experimental data.
Therefore, the main task of experimenters is conducting of new experiments
giving unambiguous conclusion on the domain of applicability and on the
domains of studied phenomena, which require its significant specification.

Calculations of probabilities of the emission of gamma-quanta in the framework
of the quasi-particle-phonon model of a nucleus and their comparison with
the experiment show that the value of $\Gamma$  is determined by the
relation of components of the quasi-particle and phonon type in the structure
of wave functions of the decaying and excited level.
In other words, the hypothesis on the independence of probability of
interaction of a gamma quantum with an excited nucleus cannot be described [4]
by some universal function, at least, lower than 3 MeV in even-even deformed
nuclei.

Similar argument but purely experimental one follows from the comparison of
the level density of various nuclei extracted [5] from intensities of two-step
cascades, on the one hand, and spectra of evaporative nucleons and primary
gamma-quanta, on the other hand.
In the first case, this parameter of nucleus demonstrates evident and abrupt
change of entropy of the nucleus at two, at the minimum, energies of its
excitation.
In the second case -- its smooth increase at the increase of the nucleus
excitation energy.
Practically, this distinction may only be related to a very small degree of
influence of the hypothesis on the independence of probability of the
emission of gamma-quanta of the given multipolarity and energy
(of an evaporative nucleon) on the structure of nucleus in the first case,
and a very strong one -- in the second case.

In other words, at present there are some data prejudicing hypothesis [1].
In this case, an experimental check of the hypothesis is obligatory with
revealing of conditions, at which its application provides for the
acceptable accuracy of experimental data.

\section{Experimental check of the Bohr-Mottelson hypothesis}

Direct extraction of the cross section $\sigma(E,U)$ in the classical experiment
(target+beam) for ordinary nuclei is practically impossible.
Nevertheless, a possibility to obtain some experimental information on the
relation $\sigma(E,U)/\sigma(E,U=0)$ for an arbitrary nucleus exists, at least,
in the (n,2$\gamma$) reaction.
The relation
\begin{equation}
P=i_{\gamma\gamma}/i_1 i_2
\end{equation}
may be determined for any level
of the nucleus under study on condition that out of two independent
experiments absolute intensities $i_{\gamma\gamma}$ of any two-step cascade
and intensities of its primary gamma transition $i_1$ and the secondary one
$i_2$ have been determined.

This value (total level population) may be presented for two arbitrary
levels $l$ and $m$ as a sum of the product of high-lying levels $l$ on the
partial cross sections of gamma-transitions $l \rightarrow m$:
$P_m=\sum P_l \times \sigma_{lm}$.
It has been determined in the form of only the cascade population $P_m-i_1$
and has been compared with different model calculations on the basis of
experimental information accumulated by now in many nuclei from the mass
region of $40 \leq A \leq 200$ for a hundred of levels almost in each of them.
The data level for $P$ may be increased abruptly in possible experiments
with more effective HPGe-detectors.

Although it is impossible to determine the value of $\sigma_{lm}$ directly
from equation (1), these experimental data due to their high sensitivity made
it possible to evaluate [5] the sign and value of the relation
$\sigma_{lm}(U<B_n)/ \sigma_{ij}(U=B_n)$.

Similar analysis may be carried out for some nuclei, in which the intensity
of two-step cascades has not been measured so far.
To do this, the data [6]
on the estimated schemes of gamma decay from the existing files are sufficient.
Unfortunately, a small value of the excitation energy overlapped by these
schemes restricts the number of nuclei, in which the observation of violation
of hypothesis [1] is possible for gamma-decay.

Such analysis has been performed [7] for $^{124}$Te.
Useful information may also be obtained for compound nuclei $^{51}$V
and $^{57}$Fe.
During the analysis one must take into account that the accuracy of
$P_m-i_1$ determination from the estimated decay schemes may not exceed the
accuracy of its determination from expression (1),
and the number of levels, for which it is possible to determine this value on
the basis of the data of type [6] is much smaller, respectively. 

Comparison of the experimental data with the calculations using various
level densities and radiative strength functions is presented in Figs. 1 and 2. 
At small excitation energies of levels good concurrence of the experiment
and any variants of the calculation is attained almost in any case due to a
very small sensitivity of the cascade population of the most low-lying
levels to variations of the calculated level densities and radiative strength
functions.
The presence of considerable discrepancy for a few levels most likely
testifies either to an erroneous determination of the level spin in the
estimated decay scheme, or to a very strong influence of the structure of
its wave function on the partial cross sections of gamma-transitions exciting
it.

\section{Conclusion}

The region of excited levels of an arbitrary nucleus with high level density
(the spacing D is smaller than the resolution FWHM of the used spectrometer)
remains so far either unexplored, or scantily explored.
Reliability of the main part of the data obtained earlier from experiment on
the level density and radiative strength functions of gamma-transitions in it
raises grave doubts, first of all, due to the absence of absolute experimental
proofs of the validity of basic hypothesis [1] of the analysis carried out here. 

Currently, there is no alternative for the explanation obtained in [5]
of the physical nature for a considerable discrepancy of the experimental data
and calculations reproducing them presented here and in [2,3].
Therefore, on the basis of the whole experimental data on intensities of
two-step cascades, with a high probability, one must not exclude a strong
influence of the nucleus structure on the level density and probabilities
of the emission of nuclear reaction products up to the neutron binding
energy or even higher energy.

\newpage
\begin{center}{References}
\end{center} \begin{flushleft} \begin{tabular}{r@{ }p{150mm}} 
$[1]$ & O. Bohr, B.R. Mottelson,{\it Nuclear Structure}, Vol. 1 
(Benjamin, NY, Amsterdam, 1969).\\
$[2]$ & H. Vonach,
{\em Proc. IAEA Advisory Group Meeting on Basic and Applied Problems
of  Nuclear Level Densities\/} (New York, 1983), INDC(USA)-092/L,
(1983) P.247\\
& B.V. Zhuravlev, Bull. Rus. Acad. Sci. Phys. 63 (1999) 123.\\ 
$[3]$ & A. Schiller et al., Nucl.  Instrum. Methods Phys.  Res. A447 (2000) 498.\\
$[4]$& V.G.  Soloviev et al., Part.  Nucl.  27(6) (1997) 1643.\\ 
$[5]$ & Sukhovoj  A.M., Khitrov V.A., 
Physics of Paricl. and Nuclei, 36(4) (2005) 359.\\
& http://www1.jinr.ru/Pepan/Pepan-index.html (in Russian)\\
$[6]$  & http://www.nndc.bnl.gov/nndc/ensdf.\\
$[7]$ &Sukhovoj  A.M., Khitrov V.A., JINR Communication E3-2007-22,
Dubna, 2007.\\
$[8]$ &W. Dilg, W.  Schantl, H.  Vonach, M.  Uhl, Nucl.  Phys. A217 (1973) 269.\\
$[9]$ &S.G. Kadmenskij, V.P. Markushev, V.I. Furman, Sov. J. Nucl. Phys. 37
(1983) 165.\\
$[10]$ & A. C. Larsen et al., Phys. Rev. C 73, 064301 (2006)\\
$[11]$ & Voinov A. et all, Phys.\ Rev. Let.,\ 2004, {\bf 93(14)} 142504-1.\\
$[12]$ & Sukhovoj A.M. et al., In:  XIII International Seminar on Interaction
of Neutrons with Nuclei,  Dubna,  May 2006, E3-2006-7, Dubna, 2006, pp. 72.
\end{tabular} 
\end{flushleft}
\newpage
\begin{figure} [htbp]
\vspace{3 cm}
\begin{center}
\leavevmode
\epsfxsize=14.5cm
\epsfbox{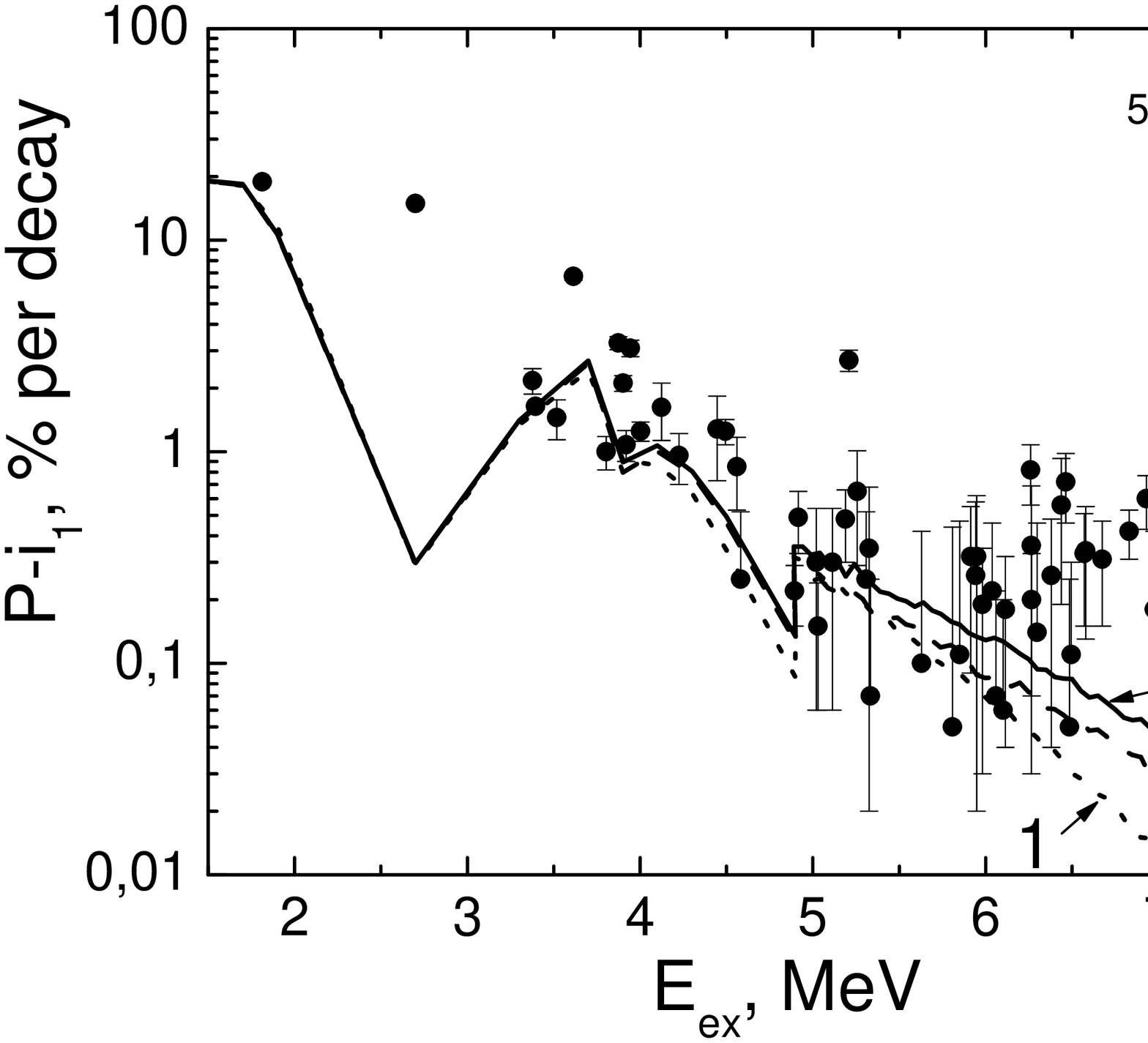}
\end{center}
\hspace{-0.8cm}\vspace{-5.cm}

{\bf Fig.~1.}~Points with errors - experimental data from [6].
Line 1 -- calculation using models [8,9]. Line 2, 3 - data from [10]
with an increase of strength functions of the low-energy E1- or M1-transitions,
only.

\end{figure}

\begin{figure}
\vspace{-9 cm}
\begin{center}
\leavevmode
\epsfxsize=12.5cm
\epsfbox{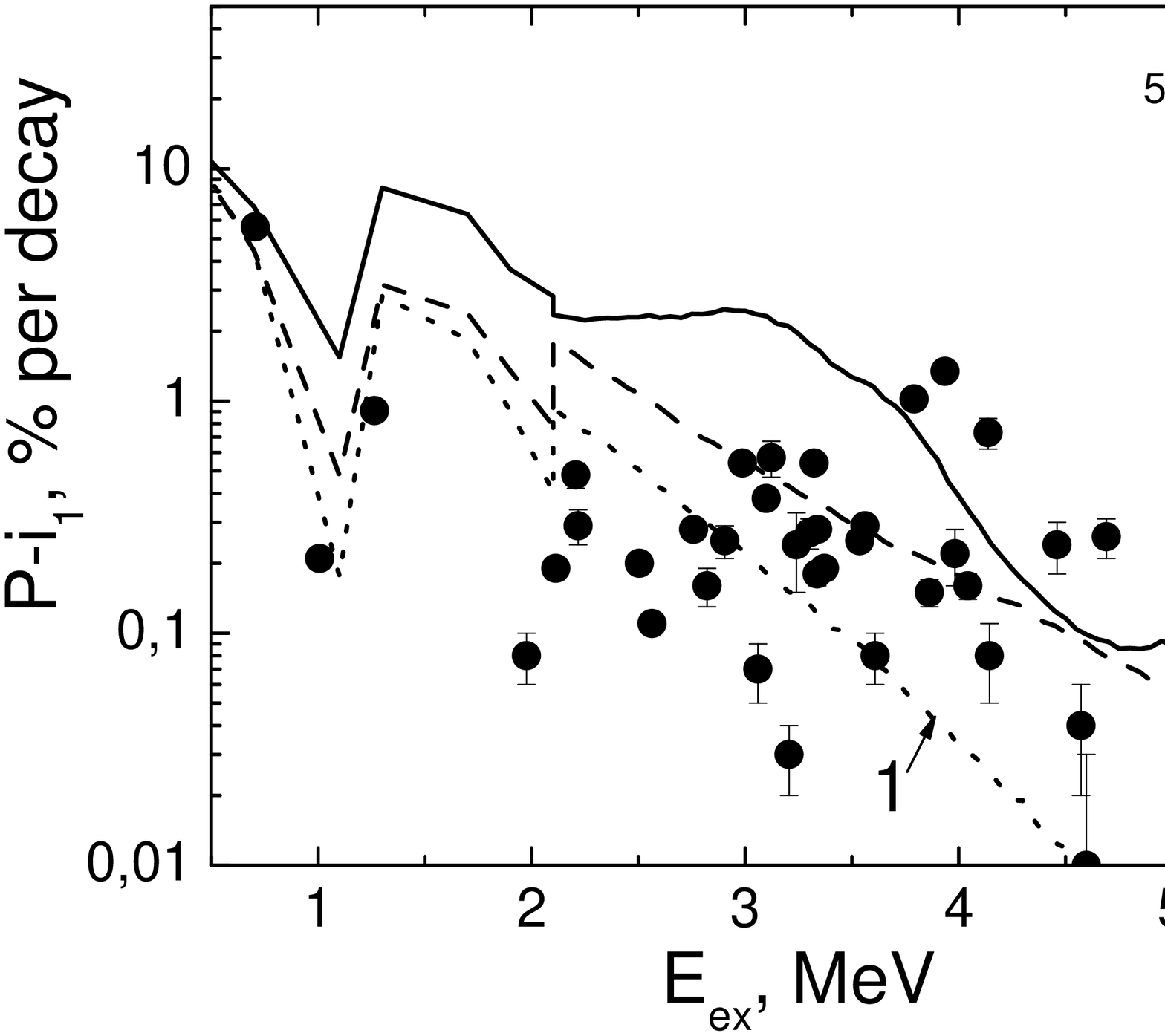}
\end{center}
\hspace{-0.8cm}\vspace{-4.5 cm}

{\bf Fig.~2.}~The same as in Fig. 1. Line 2 - data from [11]
(they do not reproduce the main part of intensities of two-step cascades).
Line 3 -- one of the variants of the level density and radiative strength
functions reproducing [12] the cascade intensity in all the 27 intervals, is used.
\end{figure}

\end{document}